\documentclass[article,twocolumn,showpacs,aps,prb,floatfix,superscriptaddress]{revtex4-1}
\usepackage{graphicx,subfigure,epsfig,verbatim,psfrag,amsmath,amssymb,color}
\usepackage{indentfirst}
\usepackage{textcomp}
\usepackage{latexsym}
\usepackage{amssymb}
\usepackage{amsmath}
\usepackage{lipsum}
\usepackage{soul}
\usepackage{xcolor}
\usepackage{color}
\usepackage{epstopdf}
\usepackage{placeins}

\usepackage{float}

\usepackage{xcolor}
\usepackage[urlcolor=blue]{hyperref}      
\hypersetup{
    colorlinks = true,                    
    citecolor = {blue},
    linkcolor = {purple},
           }
\newcommand{\be}{\begin{equation}}
\newcommand{\ee}{\end{equation}}
\newcommand{\bea}{\begin{eqnarray}}
\newcommand{\eea}{\end{eqnarray}}

\begin{document}	
	
\title{Reentrance of Topological Phase in Spin-1 Frustrated Heisenberg Chain}
\author{Yuan Yang}
\affiliation{School of Physical Sciences, University of Chinese Academy of Sciences, P. O. Box 4588, Beijing 100049, China}
\author{Shi-Ju Ran}\email[Corresponding author. Email: ] {sjran@cnu.edu.cn}
\affiliation{Department of Physics, Capital Normal University, Beijing 100048, China}
\author{Xi Chen}
\affiliation{School of Physical Sciences, University of Chinese Academy of Sciences, P. O. Box 4588, Beijing 100049, China}
\author{Zhengzhi Sun}
\affiliation{School of Physical Sciences, University of Chinese Academy of Sciences, P. O. Box 4588, Beijing 100049, China}
\author{Shou-Shu Gong}
\affiliation{Department of Physics, Beihang University, Beijing 100191, China}
\author{Zhengchuan Wang}
\email[Corresponding author. Email: ] {wangzc@ucas.ac.cn}
\affiliation{School of Physical Sciences, University of Chinese Academy of Sciences, P. O. Box 4588, Beijing 100049, China}
\author{Gang Su}
\email[Corresponding author. Email: ] {gsu@ucas.ac.cn}
\affiliation{School of Physical Sciences, University of Chinese Academy of Sciences, P. O. Box 4588, Beijing 100049, China}
\affiliation{Kavli Institute for Theoretical Sciences, and CAS Center for Excellence in Topological Quantum Computation, University of Chinese Academy of Sciences, Beijing 100190, China}
\date{\today}

\begin{abstract}
	For the Haldane phase, the magnetic field usually tends to break the symmetry and drives the system into a topologically trivial phase. Here, we report a novel reentrance of the Haldane phase at zero temperature in the spin-1 antiferromagnetic Heisenberg model on sawtooth chain. A partial Haldane phase is induced by the magnetic field, which is the combination of the Haldane state in one sublattice and a ferromagnetically ordered state in the other sublattice. Such a partial topological order is a result of the zero-temperature entropy due to quantum fluctuations caused by geometrical frustration.
\end{abstract}

\maketitle

\emph{Introduction.} --- Frustration incorporating with strong correlations usually leads to exotic quantum phenomena. A typical example in statistical physics is the reentrant phenomena, where a sequence of phase transitions, such as A-B-A-B by lowering  temperature may happen, where A and B denote ordered and disordered phases, respectively. Taking the Ising model on kagom\'e lattice with nearest and next-nearest neighboring interactions as an example, one may find that upon lowering temperature, the system passes through four phases, a paramagnetic phase, a magnetically-ordered phase, a reentrant paramagnetic phase, and a ferromagnetic phase \cite{azaria1987coexistence}. Such a reentrant phenomenon was experimentally confirmed \cite{binder1986k}, and theoretically investigated with different approaches \cite{saslow1986melting, azaria1987coexistence, debauche1991exact, diep1991exact, foster2001critical, foster2004critical, zhao2013kosterlitz}. In the past decades, systematic investigations on the partially disordered phase in both classical and quantum systems at finite temperatures were performed theoretically \cite{azaria1987coexistence, debauche1991exact, diep1991exact, quartu1997partial, santamaria1997evidence, chern2008partial, chen2011partial, ishizuka2012partial, ishizuka2013thermally, javanparast2015fluctuation} and experimentally \cite{zheng2005coexistence, zheng2006coexisting, agrestini2008incommensurate, nishiwaki2008partial, nishiwaki2011neutron, tomiyasu2012observation}. It has been widely recognized that the partially disordered phase is driven by the thermal entropy, in
which the disordered part behaves like a ``perfect'' paramagnet.  The reentrant phenomena can be viewed as the order-by-disorder effect \cite{villain1980order}, which is driven by thermal fluctuations and entropy.

Recently, the interests on the reentrant phenomena were renewed. The order-by-disorder effects were investigated theoretically \cite{jackeli2015quantum, wu2016realization, Fadi2019, gonzalez2019correlated} and experimentally \cite{savary2012order, zhitomirsky2012quantum}. For the spin-$1/2$ Heisenberg antiferromagnet on the $\sqrt{3}\times\sqrt{3}$-disordered triangular lattice \cite{gonzalez2019correlated}, a partially disordered ground state in the weakly frustrated regime was reported. The spins on the honeycomb sublattice form a $180^\circ$ N\'eel order, and spins at the hexagon center sites are in a disordered state. This work is aimed at explaining the spin-liquid behaviors of $LiZn_2Mo_3O_8$  \cite{flint2013emergent}, and shows that a partially disordered phase can be the ground state of a simple quantum isotropic Heisenberg antiferromagnet. Comparing with the classical systems, the disordered subsystem is a short range ferromagnetically correlated state instead of a paramagnet, and the mechanism is owning to the zero-point quantum fluctuations instead of the thermal ones. Furthermore, previous works show that the order-by-disorder effects are observed not just on the frustrated magnets \cite{jackeli2015quantum} but also in cold-atom platforms \cite{Fadi2019}. The quantum fluctuations might be induced by, for instance, the spin-orbit couplings \cite{Fadi2019, wu2016realization}.

\begin{figure}[tbp]
	\includegraphics[width=0.9\linewidth]{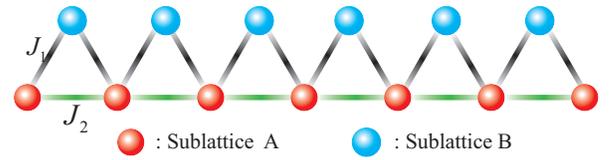}
	\caption{(Color online) Sawtooth chain lattice with two sublattice A and B. Two exchange interactions $J_1$ and $J_2$ are indicated by black and green lines, respectively.}
	\label{fig_saw}
\end{figure}

\begin{figure}[tbp]
	\includegraphics[width=1\linewidth]{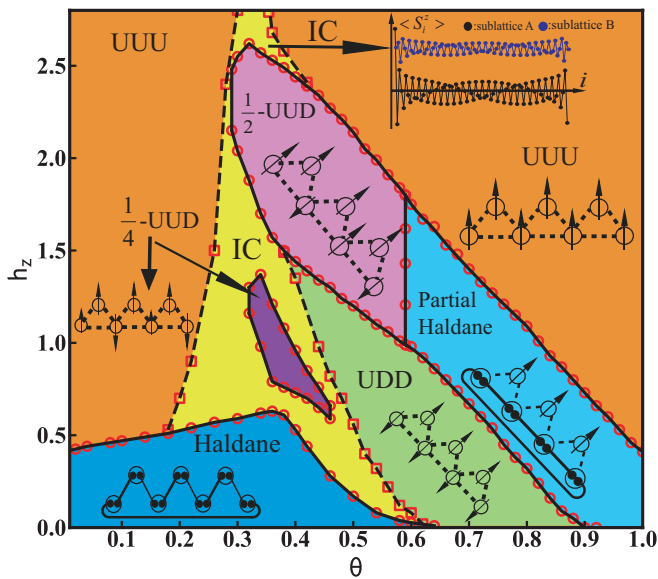}
	\caption{(Color online) The phase diagram in the $\theta$ - $h_z$ plane obtained by our finite DMRG algorithm with periodic boundary condition. The length of the sawtooth chain is $L=160$ and the truncated dimensions is $\chi= 160$. There exists a phase crossover region, which is surrounded by dashed line. There are two gapless magnetic ordered phases (UDD and UUU), and two gapped magnetic plateau phases($\frac{1}{4}$-UUD and $\frac{1}{2}$-UUD). There appear two topological ordered phases, the Haldane phase and partial Haldane phase.}
	\label{fig_phase_6}
\end{figure}

In this Letter, we study the spin-1 Heisenberg model on the sawtooth chain (Fig.~\ref{fig_saw}) by density matrix renormalization group \cite{white1992density}. Sawtooth chain is an important 1D structure that widely exists in natural materials (e.g., Cu$_2$Cl(OH)$_3$\cite{heinze2019atacamite}). The model Hamiltonian is given as
\bea \label{eq:H}
H =J_1 \sum\limits_{i} {\bf S}_{i} \cdot {\bf S}_{i+1} + J_2 \sum\limits_{\text{odd i}} {\bf S}_{i} \cdot {\bf S}_{i+2} - h_{z}\sum\limits_{i} S^{z}_{i},
\eea
where $J_1 = \cos(\frac{\pi}{2} \theta)$, $J_2 = \sin(\frac{\pi}{2} \theta)$ ($0 \leq \theta \leq 1$), and $h_z$ is the magnetic field along the $z$ direction. We report a reentrance of the Haldane phase in this model at zero temperature.
The calculational details are given in Supplementary Materials.
We establish the $\theta-h_z$ phase diagram of the system  (Fig.~\ref{fig_phase_6}), where two topologically non-trivial phases (Haldane and partial Haldane phases) are found. In the partial Haldane phase, only one sublattice of the system is in the Haldane phase  (sublattice A in Fig.~\ref{fig_saw}), and the sublattice B is in a topologically trivial phase.
Particularly for $\theta \simeq 0.6$, a reentrance of the topological Haldane phase induced by the magnetic field $h_z$ is revealed. The Haldane phase - trivial magnetic phase - partial Haldane phase transitions by increasing $h_z$ are discovered.
The partial Haldane phase appears by increasing the magnetic field.
Our work extends the quantum reentrant phenomena at zero temperature from the conventional phases (e.g., \cite{gonzalez2019correlated}) to topological phases.


\emph{Phase diagram.} --- We establish the phase diagram in the $\theta-h_z$ plane as shown in Fig. \ref{fig_phase_6}. The phase boundaries are determined by several quantities including the uniform and staggered magnetizations of the sublattices, the local magnetic moments, and the entanglement entropy measured in the middle of the system. The phase diagram consists of seven quantum phases, including two gapless magnetic phases [up-up-up phase(UUU) and up-down-down (UDD) phase with the spins pointing down (up) on the sublattice A (B)], two gapped magnetic plateau phases [up-up-down phases on the $M=\frac{1}{4}$ plateau ($\frac{1}{4}$-UUD) and $M=\frac{1}{2}$ plateau ($\frac{1}{2}$-UUD)], an incommensurate crossover (IC) region, and two topological phases (Haldane phase and partial Haldane phase). The IC and UUU are connected by crossover (implied by dash lines) with no singular behaviors \footnote{The UUU phase is not fully polarized. By further increasing $h_z$, the system will go through a $\frac{3}{4}$-plateau phase and eventually enters the fully polarized phase. See more details in the supplementary material.}.

Let us focus on the Haldane phase and partial Haldane phase. We first consider two limits $\theta=0$ and $\theta=1$. In the case of $\theta=0$, we have $J_1=1$ and $J_2=0$, and the system becomes a standard spin-1 antiferromagnetic Heisenberg chain. The ground state in this case is the well-known Haldane phase (HP) \cite{Haldane1983, Haldane1983a} with a finite gap $\Delta_0=0.41J$ \cite{white1992density, white1993numerical}. In the case of $\theta=1$, we have $J_1=0$ and $J_2=1$, the system can be regarded as Haldane chain in the sublattice A, decoupled to the free spins in the sublattice B.

\begin{figure}[tbp]
	\includegraphics[width=0.9\linewidth]{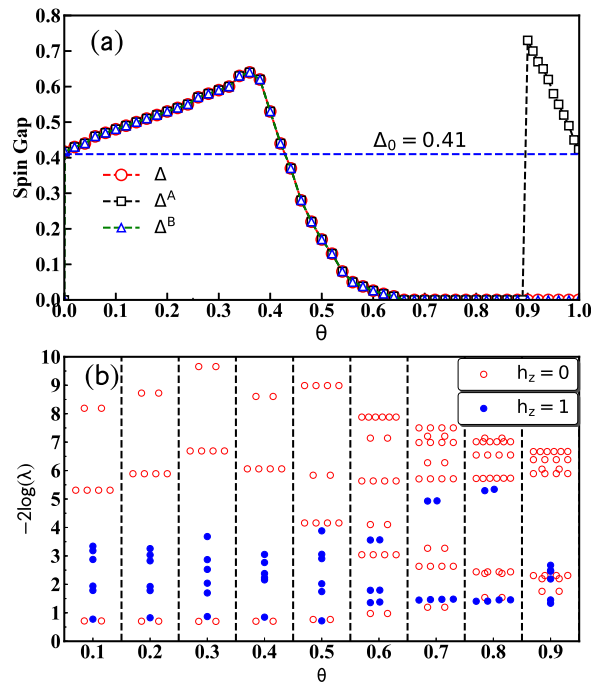}
	\caption{(Color online) (a) spin gap $\Delta$, $\Delta^A$, $\Delta^B$ versus $\theta$. (b) Entanglement spectrum versus $\theta$, where the number of dots on each level indicates its degeneracy. The red circles and the blue dots represent the cases of $h_z=0$ and $h_z=1$, respectively.}
	\label{Figure_gap_spec_v2}
\end{figure}

Fruitful physics appear at $0 < \theta < 1$. Fig. \ref{Figure_gap_spec_v2} (a) shows $\Delta^{A(B)}$ and $\Delta$ that are the spin gaps of the sublattices A (B), and that of the whole lattice.  They are determined by the width of the zero plateau of the magnetizations $M^{A(B)} = \frac{2}{N} \sum_{i\in A(B)} \langle S_i^z \rangle$ and $M=(M^A + M^B) / 2$, with $N$ the total number of sites.
For $\theta=0$, all $\Delta$ and $\Delta^{A(B)}$ are close to $\Delta_0=0.41J$, which is the gap of the standard Haldane chain. By increasing $\theta$, the spin gaps increase monotonously until $\theta \simeq 0.38$, where the gaps reach a maximum about $\Delta = \Delta^{A(B)} \simeq 0.62$. Afterwards, the spin gaps decrease rapidly and vanish at $\theta_{c1} \simeq 0.64$. For $\theta > \theta_{c1}$,  the system enters a gapless region with $\Delta = \Delta^{A(B)} = 0$ until $\theta$ reaches $\theta_{c2} \simeq 0.9$. For $\theta_{c2} < \theta < 1$, $\Delta^B$ remains zero (thus $\Delta = 0$), but $\Delta^A$ jumps to a finite value, which indicates that the sublattice A enters the Haldane phase. In this region, subsystem B is sensitive to the external field and can be polarized by a very small $h_z$, similar to a paramagnet. By increasing $\theta$ to $\theta=1$, $\Delta^A$ approaches to $\Delta^A = \Delta_0$, where the system becomes a Haldane chain decoupled with free spins.

Fig. \ref{Figure_gap_spec_v2} (b) demonstrates the entanglement spectra (ES) of the ground state. It has been suggested that ES can be used to characterize and classify exotic quantum states that are beyond the Landau symmetry-breaking paradigm \cite{li2008entanglement, Pollmann2010, fidkowski2010entanglement, yao2010entanglement}. For example, the phase transition from the Haldane phase to a topological-trivial phase can be detected by the collapse of the even-fold degeneracy structure of the ES~\cite{Pollmann2010, pollmann2012symmetry}. In our system for $h_z=0$, the ES [red hollow circles in Fig.~\ref{Figure_gap_spec_v2}(b)] verses $\theta$ shows that the even-fold degeneracy structure always appears in the whole range of $0 < \theta < 1$. This suggests that altering $J_1$ and $J_2$ will not drive the system out of the Haldane phase. This is expected since these coupling terms preserve the symmetry that protects the topological order~\cite{Pollmann2010}, which is known as the symmetry-protected topological (SPT) order~\cite{gu2009tensor}. In a SPT phase, the topological order will be robust as long as the symmetry is preserved. At $\theta = \theta_{c2} \simeq 0.9$ where the gap of sublattice A is opened, we do not see any singularity from the ES. The system should be formed by a Haldane state in sublattice A, which contributes the two-fold degeneracy, with the free spins in sublattice B. Therefore, $\theta = \theta_{c2} $ might be a topological phase transition point that separates two different kinds of string orders~\cite{Haldane1983,Haldane1983a}.

Fig.~\ref{Figure_gap_spec_v2}(b) also shows the ES versus $\theta$ at $h_z=1$ (blue solid circles). In accordance to the phase diagram, no double degeneracy of the ES is observed for $0< \theta < 0.59$ and $0.83 < \theta < 1$. These regions are formed by several phases such as magnetic plateaus and incommensurate regions which possess no topological orders. For $0.59 < \theta < 0.83$, the system is in the partial Haldane phase. Different from the $h_z=0$ case for $\theta > \theta_{c2}$, the ES shows a four-fold degeneracy. In both cases, the Haldane state in sublattice A contributes the two-fold degeneracy. For $h_z=0$, sublattice B gives a paramagnetic state, thus contributes no degeneracy to the ES. In the partial Haldane phase with $h_z > 0$, the magnetic field induces effective ferromagnetic couplings between each two nearest-neighboring spins in sublattice B. It is known that for the ground state of the ferromagnetic spin-1 chain, the two dominant values of the ES are degenerate (no degeneracy for the rest values). Therefore, the four-fold degeneracy in the partial Haldane phase ($h_z>0$) comes from both the Haldane and the effective ferromagnetic (FM) states. Within the partial Haldane phase, the state is on a $M=\frac{1}{2}$ magnetic plateau. As shown in Fig.~\ref{fig_phase_6} and Fig.~\ref{Figure_gap_spec_v2}(a), the  width of the $1/2$-plateau gives exactly the spin gap of the sublattice A. Note that the $\frac{1}{2}$ UUD also shows a $1/2$-plateau, however, the magnetic order of the sublattice A is completely different from the partial Haldane phase.

\begin{figure}[tbp]
\includegraphics[width=1\columnwidth]{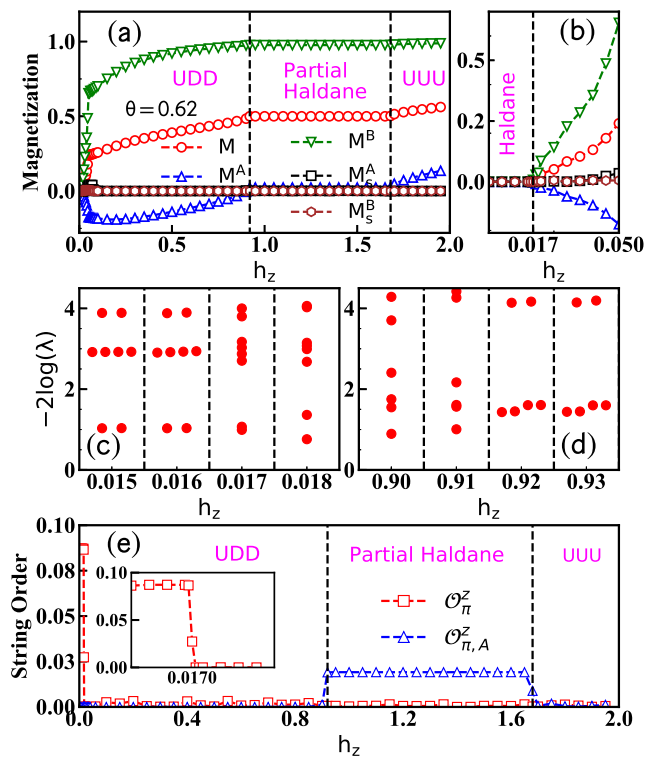}
\caption{(Color online) Detailed description of $M$, $M^A$, $M^B$, $M_s^A$ with fixed $\theta=0.62$, where $h_z$ is in the range of $[0, 2]$ in (a). (b) is the enlarged part of (a), where $h_z$ is in range of $[0, 0.05]$. The ES is illustrated in (c) and (d) in the vicinities of $h_z=0.017$ and $h_z=0.92$, respectively. (e) shows two kinds of string orders $\mathcal{O}_{\pi}^z$ and $\mathcal{O}$ at $\theta = 0.62$. The inset shows the details of $\mathcal{O}_{\pi}^z$ in the range of $0<h_z<0.035$. At $h_z=0.017$, $\mathcal{O}_{\pi}^z$ jump to zero, which is the transition point from Haldane phase to UDD magnetic ordered phase.}
\label{Figure_reen}
\end{figure}

\emph{Reentrance of topological phases.} --- Normally, topological phases including Haldane phase will be suppressed by the magnetic field. For a SPT phase, the field tends to break the symmetry and drives the system to a topologically trivial phase. In our system,  we show that the partial Haldane phase that exhibits topological order will be induced by increasing the magnetic field. The Haldane - trivial phase - partial Haldane  quantum phase transitions are observed.
Fig.~\ref{Figure_reen} presents such a reentrant behavior by showing the magnetizations and ES at $\theta = 0.62$. For $h_z < h_{c1} \simeq 0.017 $, all magnetizations are zero, and the ES shows two-fold degeneracy [Fig.~\ref{Figure_reen}(b) and (c)]. The system is in the Haldane phase. For $h_z > h_{c1}$, the system is driven to UDD phase. This is topologically trivial phase, where the spins shows long-range magnetic orders. The degeneracy of the ES is also lifted in this phase. By continuing increasing $h_z$ to $h_z > h_{c2} \simeq 0.91$, the system is driven back to a topological phase (partial Haldane phase) with two-fold degeneracy in the ES [Fig.~\ref{Figure_reen}(d)]. Until $h_z > h_{c3} \simeq 1.69$, the system enters a UUU phase.

At $\theta = 0.62$, the reentrance of the topological orders are further revealed by two kinds of string orders $\mathcal{O}_{\pi}^z$ and $\mathcal{O}_{\pi,A}^z$, which are defined as
\begin{eqnarray}
\mathcal{O}_{\pi}^z(i,j) &=& \langle \hat{S}_i^z \exp(\sum_{k=i+1}^{j-1}i\pi \hat{S}_k^z ) \hat{S}_j^z\rangle\label{Ostring}\\
\mathcal{O}_{\pi,A}^z(i,j) &=& \langle \hat{S}_i^z \exp(\sum_{k=i+1,i\in A}^{j-1,j\in A}i\pi \hat{S}_k^z ) \hat{S}_j^z\rangle \label{Ostring_a}
\end{eqnarray}
$\mathcal{O}_{\pi,A}^z$ is defined on the sublattice A, and $\mathcal{O}_{\pi}^z$ is defined on the whole lattice. In Fig. \ref{Figure_reen} (e), $\mathcal{O}_{\pi}^z$ and $\mathcal{O}_{\pi,A}^z$ are calculated in the middle of the chain by taking sufficiently large distance ($|i-j| \simeq 140$). For $h_z < 0.017$, the system is in the Haldane phase with $\mathcal{O}_{\pi}^z \simeq 0.087$ and vanishing $\mathcal{O}_{\pi,A}^z$ ($\sim O(10^{-12})$). In the partial Haldane phase for $0.91 < h_z < 1.69$, we have $\mathcal{O}_{\pi,A}^z \simeq 0.019$ and vanishing $\mathcal{O}_{\pi}^z$ ($\sim O(10^{-4})$).

Note that due to the numerical noises (particularly near the phase boundary, e.g., at $\theta=0.62$ in Fig.~\ref{Figure_reen}(d) and other results that are not shown), the four-fold degeneracy may be lifted to a double two-fold degenerate structure with a small split. This is due to the different stabilities of the degeneracies from the Haldane and effective FM parts. The degeneracy from the Haldane state is protected by the Haldane gap, thus is more stable under numerical noises. The degeneracy from the FM state is more sensitive to noises. This leads to a double two-fold structure of the entanglement spectrum.

The reentrance of the topological phases can also be understood intuitively with the effective ferromagnetic couplings in sublattice B, consistent with the four-fold degeneracy of the ES in partial Haldane phase.  In the Haldane phase, all spins participate in the Haldane state. For $0.64 < \theta < 0.9$, the system enters the UDD magnetic phase by increasing $h_z$, where the spins in the two sublattices are towards different directions. Due to the geometrical frustration, the sublattice B plays a role of a ferromagnetic background to provide an effective magnetic field $h_{inner}$ to sublattice A with $h_{inner}\propto M^B$. The $h_{inner}$ (in the opposite direction to $h_z$) dominates over both $h_z$ and the couplings $J_2$ in sublattice A, so that the spins in sublattice A cannot form a Haldane state. This explains why this region is gapless, where the $h_{inner}$ instead $h_z$ is responsible to close the Haldane gap. Until $h_z > 0.97$, the spins in sublattice B are totally polarized, and $h_z$ is strong enough to cancel with $h_{inner}$ so that the total effective field $h_{effect}=h_z-h_{inner}$ on sublattice A is beneath the Haldane gap. In this case, the spins in sublattice A forms a Haldane state, and the whole system enters the partial Haldane phase. By further increasing $h_z$, the magnetic field is sufficiently large to drive the system out of partial Haldane phase and polarizes all spins in the same direction.

\emph{Summary.} --- This work extends the reentrant phenomena to the frustrated quantum system with topological orders. We calculate the $S=1$ antiferromagnetic Heisenberg model on sawtooth lattice with different coupling strength parameterized by $\theta$. The ground-state phase diagram in a magnetic field is established, where fruitful phases are revealed. By tuning the magnetic field, an Haldane phase - trivial phase - partial Haldane phase
transitions are observed.
The partially topological-ordered phase has the coexistence of the Haldane state and topologically-trivial magnetic state in the two sublattices, respectively.

In the statistic physics, the reentrance is an entropic driven phenomenon, where the phase is stabilized by minimization of energy in combination with maximizing the entropy \cite{saslow1986melting, azaria1987coexistence, debauche1991exact, diep1991exact, foster2001critical, foster2004critical, zhao2013kosterlitz, quartu1997partial, santamaria1997evidence, chern2008partial, chen2011partial, 	ishizuka2012partial, ishizuka2013thermally, javanparast2015fluctuation}. The necessary condition for a reentrant phenomenon to occur is the existence of partial disordered phase with an ordered phase or a partial ordered phase \cite{azaria1987coexistence,debauche1991exact}. Such a mechanism was recently extended to frustrated quantum systems at zero temperature \cite{gonzalez2019correlated}, where the reentrant behavior (or the emergence of partial disorder) is a novel result of frustration and quantum fluctuations. Our work further generalizes the reentrance to topological systems, where the partially disordered phase is formed by the topological Haldane state and topologically trivial magnetic state in the two different sublattices. We expect to find more novel topological reentrant phenomena in two and higher dimensions.
\\

%

\begin{acknowledgments}
The authors acknowledge Cheng Peng, Wei Li and Yu Chen for very useful discussions. S.J.R. is supported by Beijing Natural Science Foundation (1192005  and Z180013) and Foundation of Beijing Education Committees under Grants No. KZ201810028043. S.S.G. is supported by the National Natural Science Foundation of China Grants (11834014, 11874078) and the Fundamental Research Funds for the Central Universities.
This work is also supported in part by the NSFC (Grant
No. 11834014), the National Key R\&D Program of China (Grant No. 2018FYA0305804),  the Strategetic Priority Research Program of the Chinese Academy of Sciences (Grant No. XDB28000000), and Beijing Municipal Science and Technology Commission (Grant No. Z118100004218001).
\end{acknowledgments}	

\setcounter{equation}{0}
	\setcounter{figure}{0}
	\setcounter{table}{0}
	\makeatletter
	\renewcommand{\theequation}{S\arabic{equation}}
	\renewcommand{\thefigure}{S\arabic{figure}}
	\renewcommand{\thetable}{S\arabic{table}}

\begin{appendix}


\section{Energy and entanglement entropy for $h=0$}

\begin{figure}[tbp]
\includegraphics[width=1\linewidth]{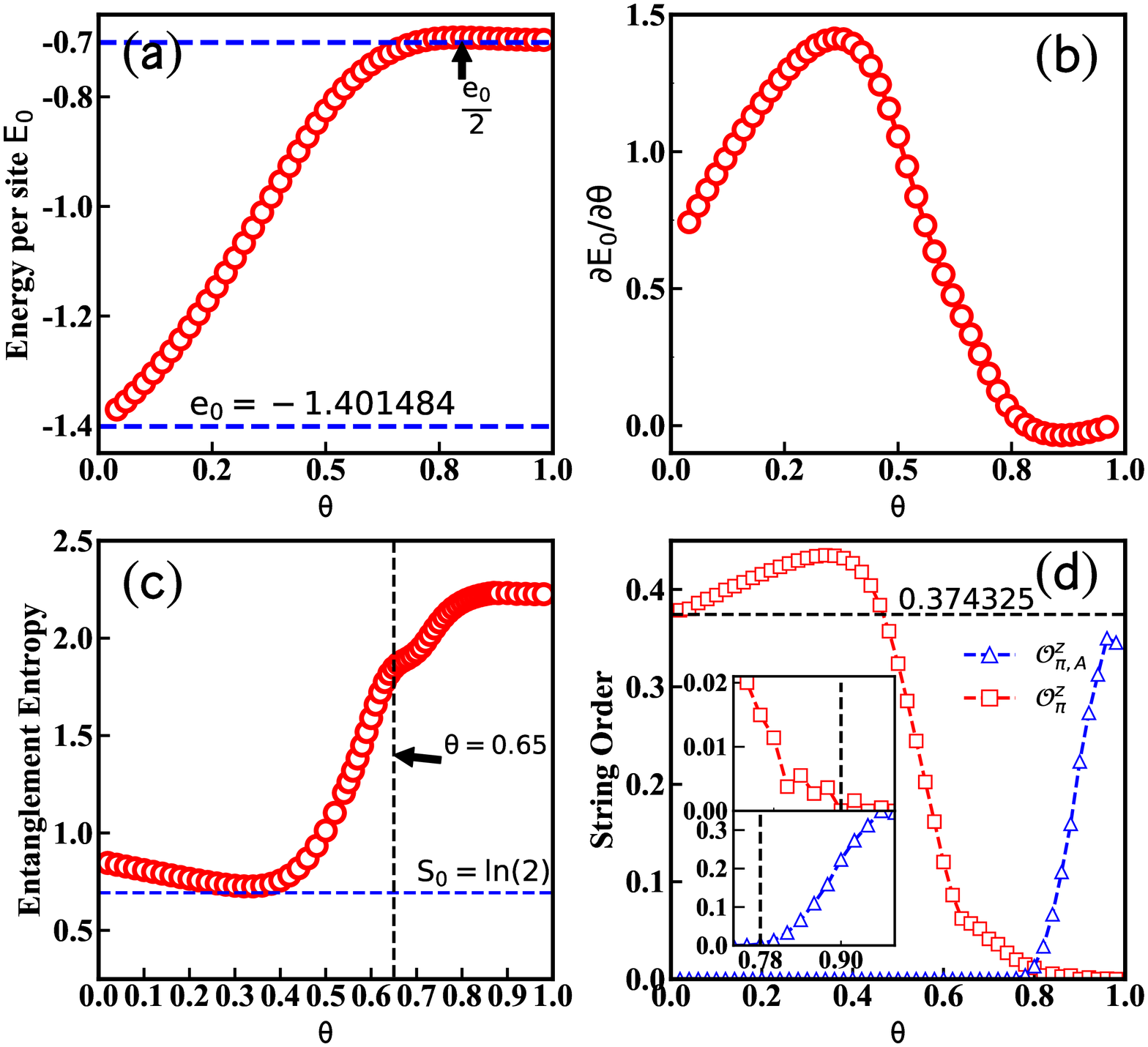}
	\caption{(Color online) (a) The variational ground-state energy per site $E_0$ versus control parameter $\theta$. In the limit of $\theta=0$, $E_0=e_0$, where $e_0$ is the ground state energy per site of the infinite antiferromagnetic $S = 1$ Heisenberg chain. In the limit of $\theta=1$, we have $E_0=e_0/2$, because the spins in the sublattice B are decoupled. The first order derivative of $E_0$ is shown in (b), the curve is continuously, i.e., there is no apparent symmetry breaking. (c) The entanglement entropy versus $\theta$, the minimum value cannot below to $ln(2)$. (d) The string orders $\mathcal{O}_{\pi}^z$ and $\mathcal{O}_{\pi,A}^z$ versus $\theta$. In the $\theta \to 0$ limit, we have  $\mathcal{O}_{\pi}^z = 0.374325$, which is consistent with the results on the standard Haldane chain. }
	\label{Figure_e_s}
\end{figure}

The Hamiltonian of our system is given by
\bea \label{eq:H}
H =J_1 \sum\limits_{i} S_{i} S_{i+1}+J_2 \sum\limits_{\text{odd j}} S_{j} S_{j+2} -h_{z}\sum\limits_{i} S^{z}_{i},
\eea
with $J_1 = \cos(\frac{\pi}{2} \theta)$, $J_2 = \sin(\frac{\pi}{2} \theta)$ ($0 < \theta < 1$), and $h_z$ the magnetic field along the $z$ direction. In Fig.\ref{Figure_e_s}, we show the ground-state energy per site $E_0$ and entanglement entropy(EE) for $h_z=0$ by density matrix renormalization group(DMRG)\cite{white1992density, schollwock2005density, schollwock2011density, mcculloch2007density} with $160$ sites and periodic condition. The entanglement entropy in the middle of the system as
\be
S=-Tr[\lambda^2 log(\lambda^2)]
\ee
with $\lambda$ the entanglement spectrum. The ground state wave function is represented as a matrix product states (MPS) \cite{fannes1989exact, Fannes1992m}.

Fig.\ref{Figure_e_s} (a) shows the ground-state energy per site $E_0$. In the limit of $\theta=0$, $E_0$ is close to the energy $e_0=—1.401484038971(4)$, which is the ground-state energy per site of the infinite antiferromagnetic $S = 1$ Heisenberg chain \cite{white1993numerical}. In the limit of $\theta=1$, $E_0$ approaches to $e_0/2$ since the spins in sublattice B are decoupled free spins and do not contribute to the total energy.

To detect if there exit any phase transitions at $h=0$, we calculate the first-order derivative of $E_0$ against $\theta$, as shown in Fig.\ref{Figure_e_s} (b). No singular behavior of $\partial E_0 / \partial\theta$ is observed. The EE of our model is shown in Fig.\ref{Figure_e_s} (c). Similarly, we do not observe any singular behavior of ground state EE against $\theta$. Only a shoulder is observed near $\theta = 0.65$ where the spin gap of the system closes (see the main text). Note that in the Haldane phase,the minimum value of entanglement entropy cannot drop below $ln(2)$ \cite{Pollmann2010}. All of our simulations imply that the system will not be driven out of the Haldane phase by $\theta$, in accordance with the data and discussions given in the manuscript.

We simulate string orders (Fig.\ref{Figure_e_s} (d)) that are defined as
\begin{eqnarray}
\mathcal{O}_{\pi}^z(i,j) &=& \langle S_i^z (exp\sum_{k=i+1}^{j-1}i\pi S_k^z )S_j^z\rangle, \label{Ostring}\\
\mathcal{O}_{\pi,A}^z(i,j) &=& \langle S_i^z (exp\sum_{k=i+1,i\in A}^{j-1,j\in A}i\pi S_k^z )S_j^z\rangle.
 \label{Ostring_a}
\end{eqnarray}
In the standard Haldane chain, one has  $\mathcal{O}_{\pi,Haldane}^z = 0.374325096$ with sufficiently large $|i-j|$ \cite{white1993numerical}. In our sawtooth model, we have $\mathcal{O}_{\pi}^z = \mathcal{O}_{\pi,Haldane}^z$ at $\theta=0$, and meanwhile $\mathcal{O}_{\pi,A}^z=0$. For $\theta > 0.9$, we have finite  $\mathcal{O}_{\pi,A}^z$ and vanishing $\mathcal{O}_{\pi}^z$. For $0.78 < \theta < 0.9$, it seems that two string orders co-exist. But our simulation suffers quite large numeric errors in this region. We for now cannot confirm whether both string orders should exist in the thermodynamic limit.


\section{Magnetic properties for $h>0$}

\begin{figure*}[tbp]
	\includegraphics[width=1\linewidth]{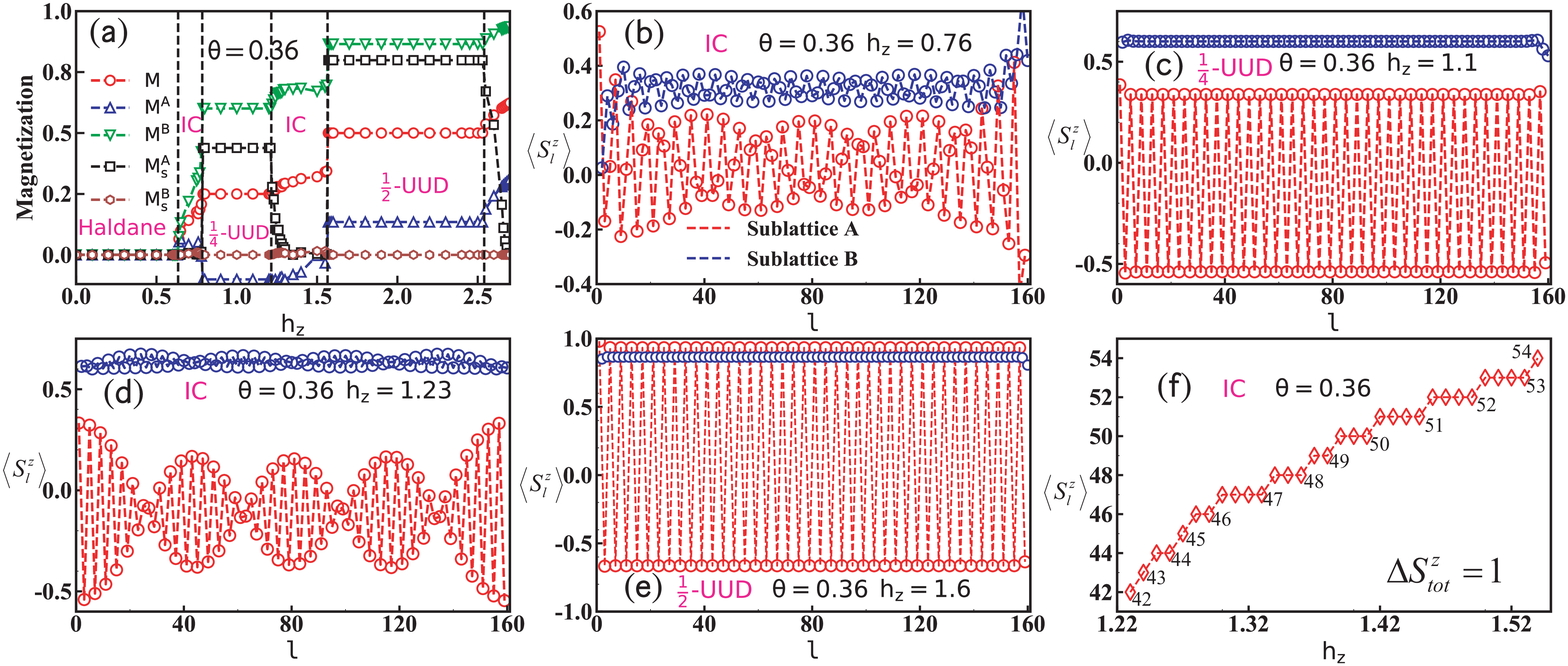}
	\caption{(Color online) (a) The magnetizations $M$, $M^{A(B)}$, $M_s^{A(B)}$ with $\theta=0.36$.  Local magnetic moment $\langle S^z_l\rangle$ in different phases at (b) $h_z=0.76$ in the IC region, (c) $h_z=1.1$ in $\frac{1}{4}$-UUD plateau phase, (d) $h_z=1.23$ in the IC region, and (e) $h_z=1.6$ in the $\frac{1}{2}$-UDD plateau phase. (f) Magnetization curve of $S^z_{tot}$ in the IC region, which increases step-wise by an integer $\Delta S^z_l=1$.}
	\label{fig_Mz_036}
\end{figure*}

We demonstrate the magnetic properties along the line $\theta=0.36$ in Fig. \ref{fig_Mz_036} (a).
The magnetizations $M$, $M^{A(B)}$, $M_s^{A(B)}$ are defined as
\begin{eqnarray}
M&=& \frac{1}{N}\sum_{i=1}^{i=N}\langle S_i^z \rangle \label{M}\\
M^{A(B)}&=& \frac{1}{N^{A(B)}}\sum_{i\in A(B)} \langle S_i^z \rangle \label{MA}\\
M_s^{A(B)}&=& \frac{1}{N^{A(B)}}\sum_{i\in A(B)}(-1)^i \langle S_i^z \rangle,
\end{eqnarray}
At $h_z=0$, the system is in the Haldane phase, where $M=M^A=M^B=M_s^A=M_s^B=0$. When $h_z=0.64$, the Haldane gap $\Delta$ is closed, and system enter into a incommensurate crossover (IC) region until $h_z=0.79$, then system get into $\frac{1}{4}$-UUD plateau phase with the width of plateau $\simeq 0.42$. By increasing $h_z$, the system enters again the IC region. For $h_z>1.57$, the system are in the $\frac{1}{2}$-UUD plateau phase and the width of this plateau is $\simeq 0.96$. At the boundary between the IC region and a plateau phase, a jump of the magnetization curve occurs, implying the existence of independent magnon excitations\cite{schulenburg2002macroscopic}.

The local magnetic moment $\langle S^z_l\rangle$ at different sites are shown in Fig. \ref{fig_Mz_036}  (b), (c), (d), and (e). At $h_z=1.1$ and $h_z=1.6$, the system is in the $\frac{1}{4}$-UUD and $\frac{1}{2}$-UUD plateau phases,respectively, as shown in (c) and (e). Spins are antiferromagnetically correlated in sublattice A, and ferromagnetically correlated in sublattice B. At $h_z=0.76$ and $h_z=1.23$, the system is in the IC region, and the local magnetic moment $\langle S^z_l\rangle$ shows a typical spin-density-wave (SDW) configuration in the real space as shown in (b) and (d). Previous studies\cite{okunishi2003magnetic, okunishi2003fractional, hikihara2008vector, hikihara2010magnetic} on frustrated spin ladders showed that the total magnetization $S_{tot}^z$ changes by an integer in the SDW phase, i.e. $\Delta S_{tot}^z=integer$, where $S_{tot}^z=\sum_{l} \langle S_l^z\rangle$. We calculate the $S^z_{tot}$ magnetization curve in the range of $1.23\leq h_z \leq 1.54$ as shown in (d), showing $\Delta S^z_l=1$. This means that the low-lying excitation in the IC region corresponds to a single spin flip.

\begin{figure}[tbp]
	\includegraphics[width=1\linewidth]{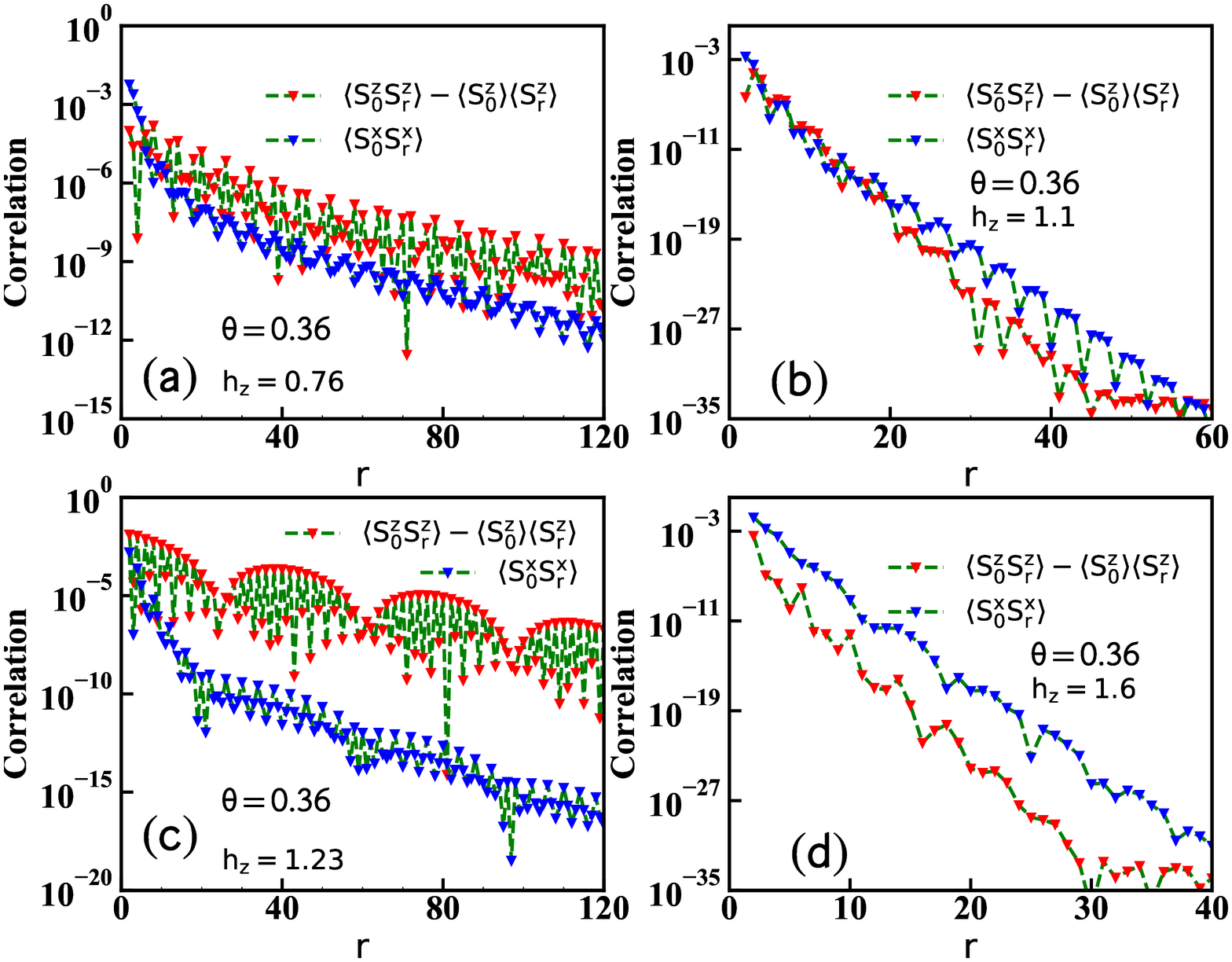}
	\caption{(Color online) Spatial dependence of the fluctuations $\langle S_0^z S_r^z\rangle -\langle S_0^z\rangle\langle S_r^z\rangle$ and $\langle S_0^x S_r^x\rangle$ with fixed $\theta=0.36$ and (a) $h_z=0.76$, (b) $h_z=1.1$, (c) $h_z=1.23$, and (d) $h_z=1.6$. When $h_z=0.76$ and $h_z=1.23$, l$\langle S_0^z S_r^z\rangle -\langle S_0^z\rangle\langle S_r^z\rangle$ is dominant. When $h_z=1.1$ and $1.23$, both $\langle S_0^z S_r^z\rangle -\langle S_0^z\rangle\langle S_r^z\rangle$ and $\langle S_0^z S_r^x\rangle$ decay exponentially. }
    \label{fig_t036_corre}
\end{figure}

\begin{figure}[tbp]
	\includegraphics[width=0.9\linewidth]{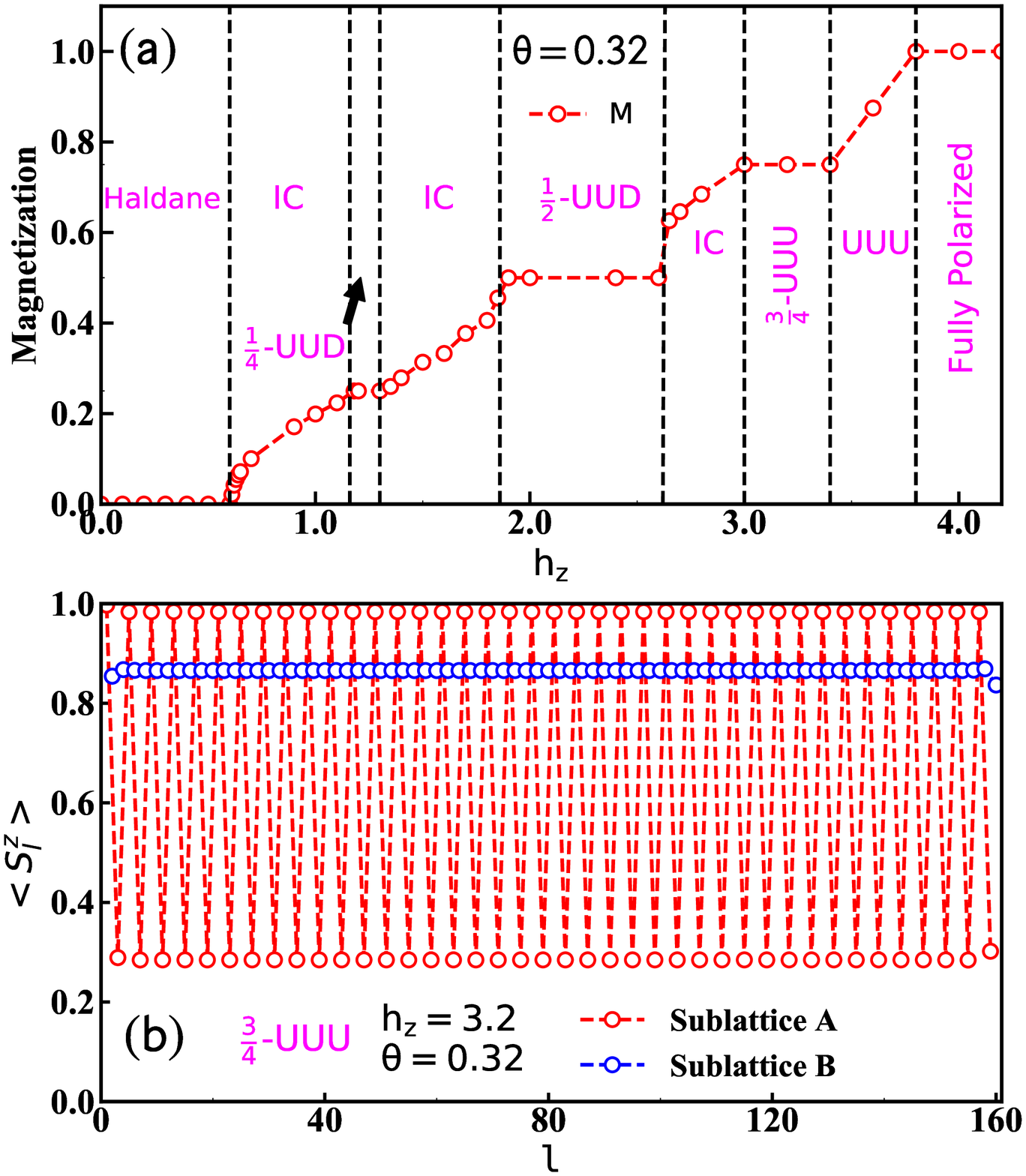}
	\caption{(Color online)(a) The magnetization per site $M$ for $\theta=0.32$. (b) Local magnetic moment $\langle S^z_l\rangle$ for $\theta=0.32$ and $h_z=3.2$ within the $\frac{3}{4}$-UUU phase.}
	\label{fig_Mz_032}
\end{figure}

When $h_z=1.1$ and $h_z=1.6$, system is stabilized in the $M=\frac{1}{4}$ and $M=\frac{1}{2}$ plateau phases, The spin configurations are shown by the local magnetizations in Fig. \ref{fig_Mz_036} (c) and (e). Both of the two phases exhibit long range order (LRO) with a same up-up-down structure. Spins are antiferromagnetically correlated in sublattice A and ferromagnetically correlated in sublattice B. We also calculate the spin fluctuations $\langle S_0^z S_r^z\rangle -\langle S_0^z\rangle\langle S_r^z\rangle$ and $\langle S_0^x S_r^x\rangle$. As shown in Fig. \ref{fig_t036_corre} (b) and (d), both decay exponentially since the states are gapped.

For $h_z=0.76$ and $h_z=1.23$, it is obvious that $\langle S^z_l \rangle$ breaks the translational invariance and exhibits a typical SDW. In general, the dominant spin correlation in a SDW are the component parallel to the field\cite{chen2013ground}. To confirm whether a SDW emerges in the IC region, we compute $\langle S_0^z S_r^z\rangle -\langle S_0^z\rangle\langle S_r^z\rangle$ and $\langle S_0^x S_r^x\rangle$ as shown in Fig.\ref{fig_t036_corre} (a) and (c) at $h_z=0.76$ and $h_z=1.23$, respectively. Both spin fluctuations show the incommensurate feature. Furthermore, $\langle S_0^z S_r^z\rangle -\langle S_0^z\rangle\langle S_r^z\rangle$ dominates over other fluctuations. Combined with property of $\Delta S^z_{tot} = 1$ as shown in Fig.\ref{fig_Mz_036}(f), evidently the IC region is a kind of SDW phase.

Besides the phases shown in the phase diagram in the main text, there are additionally two phases with relatively large $h_z$, which are $\frac{3}{4}$-plateau phase with an up-up-up structure (denoted as $\frac{3}{4}$-UUU) and the fully polarized phase. Fig. \ref{fig_Mz_032} (a) shows the magnetization per site at $\theta=0.32$, where the system goes through the $\frac{3}{4}$-UUU and fully polarized phases for about $h_z > 3$. Fig. \ref{fig_Mz_032} (b) shows the local magnetic moment $\langle S^z_l\rangle$ within the $\frac{3}{4}$-UUU phase ($h_z = 3.2$). The spins in the sublattice A are arranged ferrimagnetically, and the spins in the sublattice B are uniformly polarized with an unsaturated  magnetization.

\section{Magnetizations and entanglement related to the partial Haldane phase}

\begin{figure}[tbp]
	\includegraphics[width=1\linewidth]{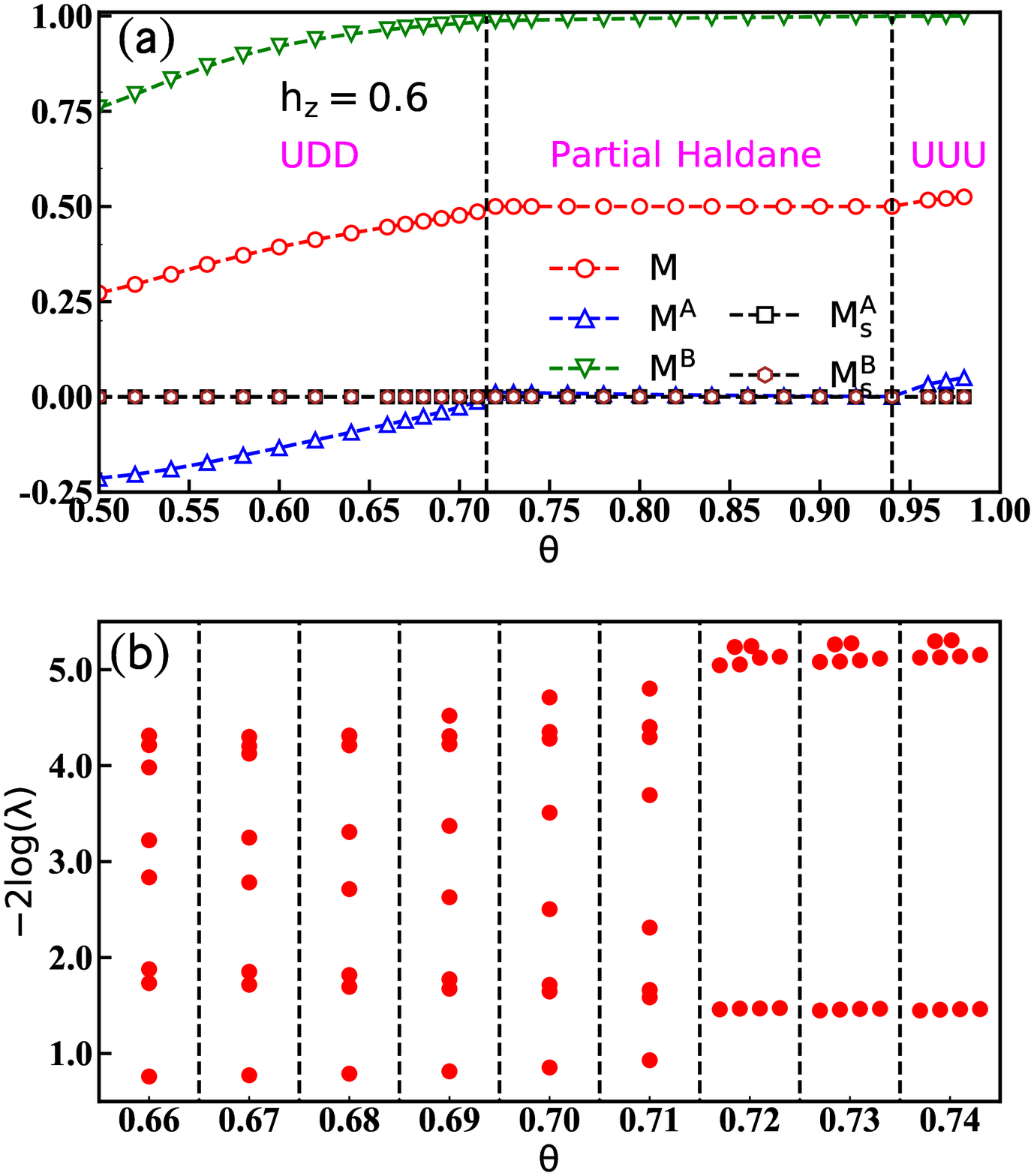}
	\caption{(Color online) (a) The magnetizations $M$, $M^A$, $M^B$, $M_s^A$ versus $\theta$ with $h_z=0.6$. (b) The entanglement spectrum (logarithmic plot). When $\theta<0.72$, the spectrum is not degenerate; when $\theta>0.72$, the whole spectrum shows even-fold degeneracies.}
	\label{fig_hz06}
\end{figure}

\begin{figure}[tbp]
	\includegraphics[width=1\columnwidth]{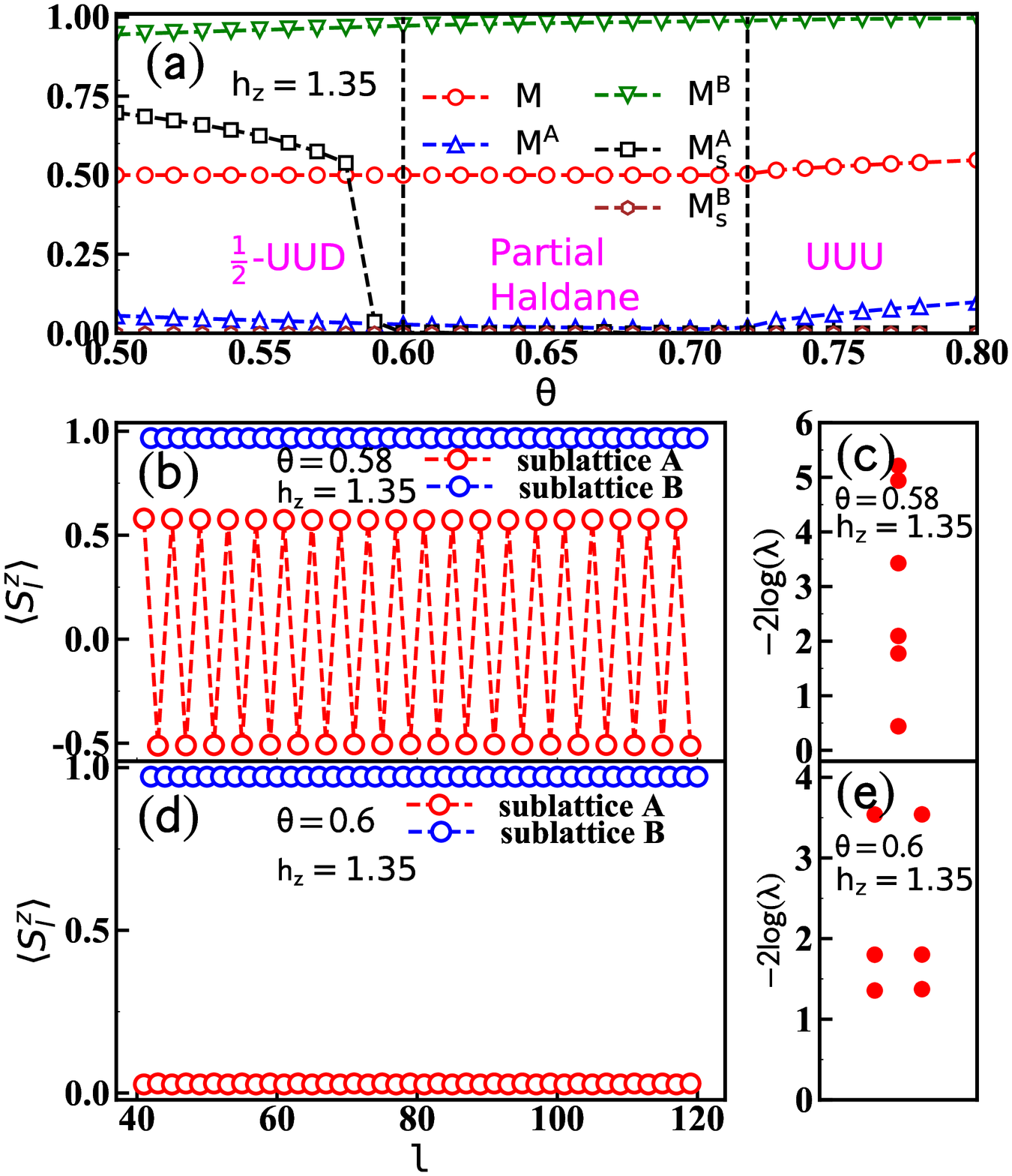}
	\caption{(Color online) (a) The magnetizations $M$, $M^A$, $M^B$, $M_s^A$ versus $\theta$ at $h_z=1.35$. The local magnetic momentum $\langle S_i^z\rangle$ at different sites are illustrated in (b) and (d) with $\theta=0.58$ and $\theta=0.6$. The entanglement spectrum (logarithmic plot) is showed in (c) and (e).}
	\label{fig_detail_2}
\end{figure}

\begin{figure}[tbp]
	\includegraphics[width=1\columnwidth]{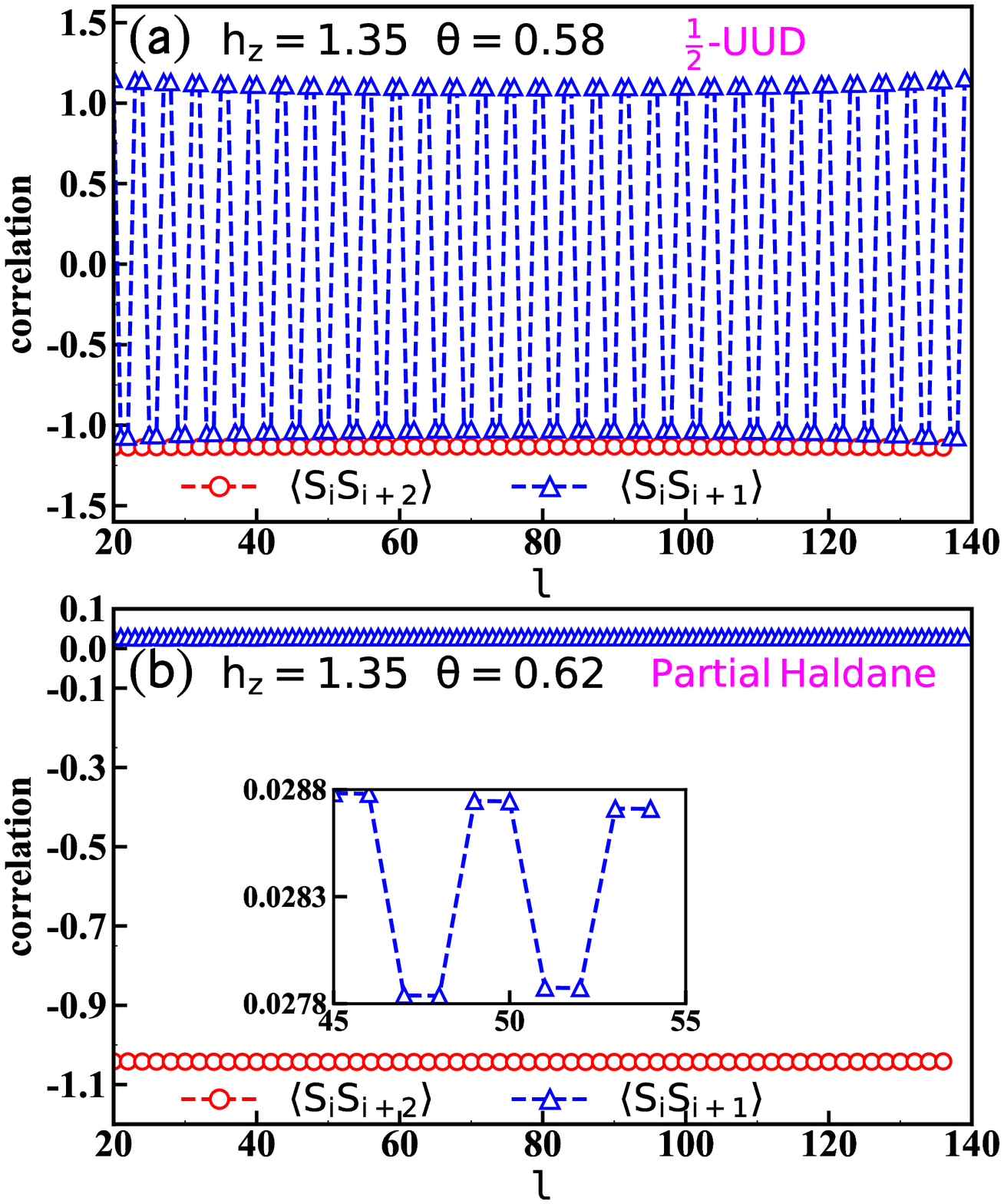}
	\caption{(Color online) At $h_z=1.35$, the nearest-neighbor spin correlation $\langle S_iS_{i+1}\rangle$ 		and next nearest-neighbor spin correlation $\langle S_iS_{i+2}\rangle$ of (a) the $\frac{1}{2}$UUD plateau phase with $\theta=0.58$ and (b) partial Haldane phase with $\theta=0.72$. The inset of (b) shows that the errors that break the translational invariance of the Haldane phase in sublattice A is only $O(10^{-4})$.}
	\label{fig_hz135_Eb}
\end{figure}

Fig. \ref{fig_hz06} shows the magnetizations and entanglement $h_z=0.6$, where the system goes through the UDD, partial Haldane, and UUU phases. In Fig.\ref{fig_hz06} (a), we have $M^A<0$, $M^B>0$ and $M_s^{A(B)}=0$ when $\theta<0.72$, which indicates that the spins is ferromagnetic correlated within both sublattices A and B, but are antiferromagnetically arranged between A and B. The spins point down in A and point up in B. When $\theta\geq=0.72$, we have $M^A=M_s^A=0$ suggesting that the sublattice A enters the Haldane phase with vanishing magnetizations. Meanwhile, we have $M^B=1$ showing that the spins in sublattice B are fully polarized. So the uniform magnetization of the whole system $M$ equals to $0.5$, giving a $1/2$ magnetic plateau.

Fig.\ref{fig_hz06} (b) shows the entanglement spectrum(logarithmic plot) versus $\theta$. For $\theta<0.72$, no degeneracy is observed in the spectrum. For $\theta>0.72$, the whole spectrum shows even-fold degeneracies, suggesting that the sublattice A enters a Haldane phase. Interestingly, the dominant values of the spectrum are four-fold degenerate. This is different from the standard Haldane phase where the dominant part shows two-fold degeneracy. The four-fold degeneracy is a combined result of the Haldne nature of sublattice A and the effective ferromagnetic interactions in sublattice B. The two sublattices, though possessing different nature, both contribute two-fold degeneracy, resulting in a four-fold degeneracy in the entanglement spectrum of the total ground state.

Both the $\frac{1}{2}$-UUD and partial Haldane phases are on the $\frac{1}{2}$ magnetic plateau. However, the properties of these two phases are completely different.  Fig.\ref{fig_detail_2} (a) shows the magnetizations at $h_z=1.35$. For $\theta<0.6,$ the system is in $\frac{1}{2}$-UUD with$M=0.5$, and both $M^A$ and $M_s^A$ are finite. For $\theta=0.6$, both $M_s^A$ and $M^A$ vanish to zero at the same time. Fig. \ref{fig_detail_2} (b) and (d) give the local magnetic momentums in the two phases, respectively. Usually a plateau magnetization is a commensurate, classical state stabilized by quantum fluctuations. This is well reflected in Fig.\ref{fig_detail_2} (b) where the system is in $\frac{1}{2}$-UUD phase and the local magnetic momentums in the sublattice A exhibit a classical antiferromagnetic order. In contrast for $\theta\geq 0.6$ as showed in (d), the $\frac{1}{2}$ plateau is formed by the magnetically disordered Haldane state in A and the classical polarized state in B. Besides the local magnetic momentum, the $\frac{1}{2}$-UUD and partial Haldane phases can be distinguished by the entanglement spectrum as shown in Fig. \ref{fig_detail_2} (c) and (e). The ES is not degenerate at $h_z=1.35$ and $\theta=0.58$, while it shows double degeneracy at $\theta=0.6$.

In Fig. \ref{fig_hz135_Eb}, we show the nearest-neighbor (NN) spin correlation $\langle S_iS_{i+1}\rangle$ and next-nearest-neighbor (NNN) spin correlation $\langle S_iS_{i+2}\rangle$. Note $\langle S_iS_{i+2}\rangle$ is in fact the NN spin correlation on sublattice A ($J_2$ couplings). As shown in Fig. \ref{fig_hz135_Eb} (a), when system is in $\frac{1}{2}$-UUD plateau phase, obviously the translational invariance is broken. The correlation is consisted with the pattern of local magnetic momentum shown in Fig.\ref{fig_detail_2} (b).

As the local magnetic moment is equal to zero on sublattice A in the partial Haldane phase, it is also possible that the spins form valance bonds. If this is true, the translational invariance should be broken. To exclude this case, we show $\langle S_iS_{i+1}\rangle$ and $\langle S_iS_{i+2}\rangle$ in the partial Haldane phase in Fig. \ref{fig_hz135_Eb} (b). Both the $\langle S_iS_{i+1}\rangle$ and $\langle S_iS_{i+2}\rangle$ are constant, meaning the translational invariance is kept. The inset of (b) shows that the difference of $\langle S_iS_{i+1}\rangle$ is around $O(10^{-4})$ due to computational errors. The translational symmetry is well kept and such state is not a valence-bond state.

\end{appendix}
%

\end{document}